\pdfoutput=1  
\documentclass[aps,prd,superscriptaddress,floatfix,nofootinbib,preprintnumbers,eqsecnum,onecolumn]{revtex4-1}

\RequirePackage{amsmath}
\RequirePackage{amssymb}
\RequirePackage{epsfig}
\RequirePackage{graphicx}
\RequirePackage{layout}
\RequirePackage{setspace}
\RequirePackage{enumitem}
\RequirePackage{soul}
\RequirePackage{eurosym}
\RequirePackage{url}
\RequirePackage{verbatim}
\RequirePackage{moreverb}
\RequirePackage{listings}
\RequirePackage{amsthm}
\RequirePackage{mathrsfs}
\RequirePackage{cancel}
\RequirePackage{makeidx}
\RequirePackage{multirow}
\RequirePackage{stmaryrd}
\RequirePackage{xcolor}
\RequirePackage{lmodern}
\usepackage[normalem]{ulem}

\usepackage{tikzsymbols}
\usepackage{natbib}
\usepackage{float}

\usepackage{tikz,xcolor,hyperref}

\definecolor{lime}{HTML}{A6CE39}
\DeclareRobustCommand{\orcidicon}{
	\begin{tikzpicture}
	\draw[lime, fill=lime] (0,0) 
	circle [radius=0.2] 
	node[white] {{\fontfamily{qag}\selectfont \tiny ID}};
	\draw[white, fill=white] (-0.0625,0.095) 
	circle [radius=0.007];
	\end{tikzpicture}
	\hspace{-2mm}
}

\foreach \x in {A, ..., Z}{\expandafter\xdef\csname orcid\x\endcsname{\noexpand\href{https://orcid.org/\csname orcidauthor\x\endcsname}
			{\noexpand\orcidicon}}
}

\renewcommand{\bar}[1]{\overline{\mathstrut #1}}
\renewcommand{\vec}[1]{\overrightarrow{\mathstrut #1}}

\newcommand{\be}{\begin{equation}}
\newcommand{\ee}{\end{equation}}
\newcommand{\bea}{\begin{eqnarray}}
\newcommand{\eea}{\end{eqnarray}}

\newcommand{\ag}[1]{\textcolor{blue}{[Anish: #1]}}
\newcommand{\lucien}[1]{\textcolor{cyan}{#1}}

\begin{document}




\title{Signatures of Non-thermal Dark Matter with Kination and Early Matter Domination: \\ \it{Gravitational Waves versus Laboratory Searches}}



\author{Anish Ghoshal\orcidA{}}
\email{anish.ghoshal@fuw.edu.pl}
\affiliation{Institute of Theoretical Physics, Faculty of Physics, University of Warsaw, ul. Pasteura 5, 02-093 Warsaw, Poland}

\author{Lucien Heurtier\orcidB{}}
\email{lucien.heurtier@durham.ac.uk}
\affiliation{Institute for Particle Physics Phenomenology, Durham University, South Road, Durham, U.K.}

\author{Arnab Paul\orcidC{}}
\email{arnabpaul9292@gmail.com}
\affiliation{School of Physical Sciences, Indian Association for the Cultivation of Science, Kolkata-700032, India}

\preprint{IPPP/22/54}

\begin{abstract}
The non-thermal production of dark matter (DM) usually requires very tiny couplings of the dark sector with the visible sector and therefore is notoriously challenging to hunt in laboratory experiments. Here we propose a novel pathway to test such a production in the context of a non-standard cosmological history, using both gravitational wave (GW) and laboratory searches. We investigate the formation of DM from the decay of a scalar field  that we dub as the reheaton, as it also reheats the Universe when it decays. We consider the possibility that the Universe undergoes a phase 
with \textit{kination-like} stiff equation-of-state ($w_{\rm kin}>1/3$) before the reheaton dominates the energy density of the Universe and eventually decays into Standard Model and DM particles. We then study how first-order tensor perturbations generated during inflation, the amplitude of which may get amplified during the kination era and lead to detectable GW signals. Demanding that the reheaton produces the observed DM relic density, we show that the reheaton's lifetime and branching fractions are dictated by the cosmological scenario. In particular, we show that it is long-lived and can be searched by various experiments such as DUNE, FASER, FASER-II, MATHUSLA, SHiP, etc. We also identify the parameter space which leads to complementary observables for GW detectors such as LISA and u-DECIGO. In particular we find that a kination-like period with an equation-of-state parameter $w_{\rm kin}\approx 0.5$ and a reheaton mass $\mathcal O(0.5-5)$ GeV and a DM mass of $\mathcal O (10-100)$ keV may lead to sizeable imprints in both kinds of searches. 
\end{abstract}

\maketitle

\section{Introduction}
\label{sec:intro}

The cosmic inflation which resolves the flatness and horizon problem and seeds the initial density fluctuations for large-scale structure formation~\cite{Brout:1977ix,Sato:1980yn,Guth:1980zm,Linde:1981mu,Starobinsky:1982ee} predicts
tiny fluctuations in the Cosmic Microwave Background (CMB) measurements \cite{Planck:2018vyg}. After its discovery, the content of our present Universe in Dark Matter (DM), Dark Energy (DE) and radiation is now well established in what is known as the $\Lambda$CDM model \cite{mambrini2021particles}. In addition, the improving measurements of the scalar perturbation modes, together with the most recent limits on the presence of tensor modes in the CMB, help narrowing down the class of inflation models which could explain the incredible homogeneity and flatness of the Universe. Nevertheless, the history of the Universe from the end of cosmic inflation to the hot big bang phase of the cosmological history remains up to now free of any experimental constraints. As a consequence, the way the metric perturbation modes evolve after their production during inflation, until the present time, is unknown. The consequences of this complete black out regarding our Universe history is twofold: $(i)$ We are unable to predict with certainty the scale of inflation and $(ii)$ the number of $e$-folds of cosmic inflation, which is essential to constrain cosmic inflation models from the CMB measurement, is a free parameter of the theory.

In the vanilla $\Lambda$CDM model, it is generally assumed that the cosmic inflation era is followed immediately by the radiation dominated (RD) era, also known as the hot big bang phase of the cosmological history. In this very simplistic case, it is expected that the spectrum of GWs that was produced during inflation remained frozen until perturbation modes start growing linearly with the expansion during the late Matter-Domination (MD) era. Since inflation is measured to produce a nearly scale invariant spectrum of first-order tensor perturbations that is relatively feeble as compared to the sensitivity of present and future GW detectors, it is expected that a Universe exclusively dominated by radiation and matter after inflation would not lead to any measurable GW signal in the near future. However, we would like to highlight that the Universe can only become radiation dominated at the end of inflation under very conservative assumptions. Indeed, to release all of its energy density right after it exits the phase of {\em slow roll}, the inflaton must decay immediately into ordinary radiation. Such a fast decay of the inflaton field requires the existence of large interaction terms between the inflaton field and Standard Model (SM) fields. However, sizeable interactions of the inflationary sector with the SM are not motivated by any strong theoretical argument, they were also shown to substantially affect the inflationary dynamics \cite{Buchmuller:2014pla,Buchmuller:2015oma,Argurio:2017joe,Heurtier:2019eou} or the stability of the SM Higgs boson \cite{Enqvist:2016mqj, Kost:2021rbi}. 
Furthermore, in order to decay efficiently after inflation ends, the inflaton also needs to oscillate around the minimum of its potential, such that its coherent oscillations quickly get damped by through SM particle production. This relies on the idea that the inflation potential minimum stands relatively close in field space from the point where inflation ends. However, numerous scalar potentials can be used to realize cosmic inflation which do not have a minimum or whose minimum is far away from the location in field space where inflation ends. This is for instance the case of quintessential inflation models  \cite{Akrami:2017cir, Dimopoulos:2001ix, Bettoni:2021qfs}, or more simply non-oscillatory inflation models \cite{Ellis:2020krl}, in which the inflaton keeps rolling along its potential for a long time after inflation ends. In such cases, the production of SM particles is more difficult to achieve and can typically be realized through gravitational particle production~\cite{Ford:1986sy,Chun:2009yu} or other reheating mechanisms, just to name a few, instant preheating \cite{Dimopoulos:2017tud}, curvaton reheating \cite{Feng:2002nb}, Ricci reheating \cite{Opferkuch:2019zbd,Bettoni:2021zhq}. The  inflation sector thus only transfers a fraction $\eta$ of its energy density when SM particles are produced. The Universe therefore undergoes a phase of kination, where the kinetic energy of the inflaton scalar field is the main source of energy in the Universe and decreases quickly with expansion as $\rho_\phi\sim a^{-6}$ before radiation starts dominating and the hot big bang phase starts.

\begin{figure}
    \centering
    \includegraphics[width=\linewidth]{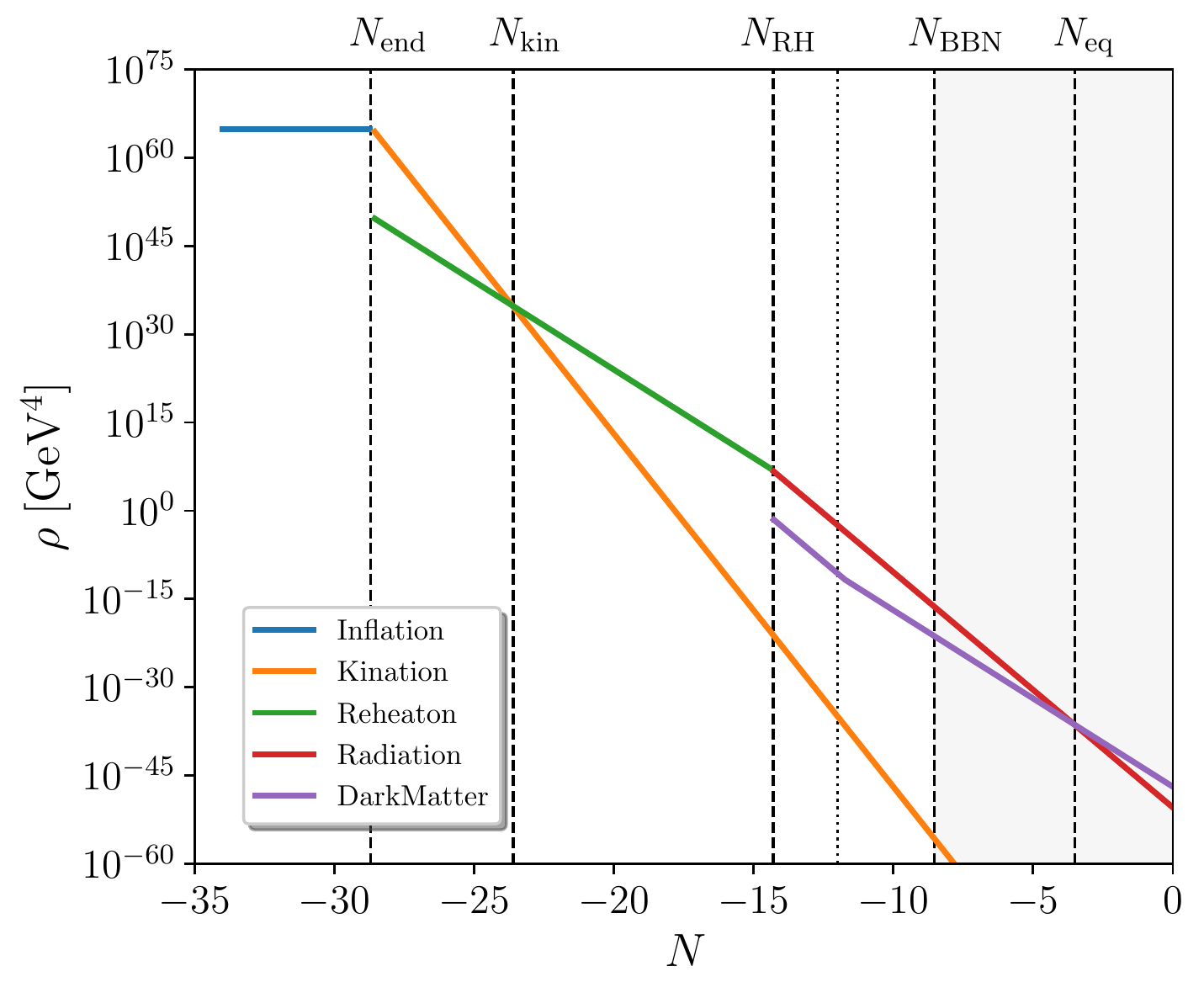}
    \caption{\it \label{fig:drawing} \footnotesize Evolution of the different components present in the Universe along the cosmological history. The vertical dotted line stands for the time where DM becomes non-relativistic.The parameters used in the figure are $\eta=10^{-15}$, $m_S=1\,\mathrm{GeV}$, $m_{\rm DM}=1\,\mathrm{MeV}$, $\Gamma_S=10^{-15}\,\mathrm{GeV}$, and $w_{\rm kin}=1$. In the $x$-axis, $N=\ln(a)$ stands for the number of $e$-folds before present time.}
    
\end{figure}

In this paper we will consider the more general possibility that the end of inflation is not continued right away by the hot big bang phase, but instead is followed by 
a phase featuring a stiff equation of state, corresponding to an equation of state parameter larger than the one of radiation that we will denote as $w_{\rm kin}>1/3$. In what follows, we will refer to this period as being \textit{'kination-like'} for simplicity. For completeness, we also envision that the inflaton may not produce SM particles directly but may instead produce a metastable moduli (called {\em reheaton}) that will later on reheat the Universe and produce dark-matter particles out of equilibrium. This transfer of energy could correspond to a tiny gravitational particle production, but can also simply arise from a slight transfer of kinetic energy from the inflaton oscillations to the oscillations of a transfer direction in field space. This happens typically in supergravity models when spectator scalar fields may have Hubble-size masses during inflation but small masses in the vacuum \cite{Heurtier:2022rhf,Heurtier:2015ima,Gonzalo:2016gey, Argurio:2017joe} and therefore may start ocsillating at the end of inflation. For simplicity, we will assume that the oscillations of the reheaton can be described by a perfect fluid with constant equation of state parameter $w_S$. Typically, if the reheaton oscillates around quadratic potential, this fluid behaves like cold matter with equation of state $w_S=0$, which would lead to a period of early matter domination. In that way, the History of the Universe, also depicted in Fig.~\ref{fig:drawing} can be described by four major phases:
\bea
&\mathrm{Inflation}&\nonumber\\
&\downarrow&\nonumber\\
&\mathrm{Kination\ /\ Stiff\ Equation\ of\ State}&\nonumber\\
&\downarrow&\nonumber\\
&\mathrm{Early\ Matter\ Domination \,(EMD)}&\nonumber\\
&\downarrow&\nonumber\\
&\mathrm{SM\ Plasma} + {\rm Dark\   Matter}&\nonumber
\eea
In a thorough study, the authors of Ref.~ \cite{Gouttenoire:2021jhk} showed recently that a kination-like period can lead to a growth of perturbation modes of the metric at large frequencies that could be detectable by gravitational wave detectors in the near future ~\cite{ Giovannini:1998bp, Giovannini:2009kg, Riazuelo:2000fc, Sahni:2001qp, Seto:2003kc, Tashiro:2003qp, Nakayama:2008ip, Nakayama:2008wy, Durrer:2011bi, Kuroyanagi:2011fy, Kuroyanagi:2018csn, Jinno:2012xb, Lasky:2015lej, Li:2016mmc, Saikawa:2018rcs, Caldwell:2018giq, Bernal:2019lpc, Figueroa:2019paj, DEramo:2019tit,Li:2013nal, Odintsov:2021kup, Odintsov:2021urx, Li:2016mmc, Li:2021htg,Dimopoulos:2022mce,Co:2021lkc,Gouttenoire:2021jhk,Vagnozzi:2020gtf,Benetti:2021uea,Hook:2020phx}. We will therefore explore to which extent such a possibility remains promising for this generic scenario and exhibit regions of the parameter space which will be probed by future GW detectors.

Besides the unknown pre-BBN history of the universe, the origin and composition of dark matter (DM) in the Universe remains a big mystery in modern particle physics and cosmology~\cite{Jungman:1995df,Bertone:2004pz,Feng:2010gw}. Despite great experimental efforts over the last 30 years, the simplest models of a dark matter particle that freezes out from the SM plasma leading to the famous ``WIMP miracle'' ~\cite{Lee:1977ua,Scherrer:1985zt,Srednicki:1988ce,Gondolo:1990dk} were not detected experimentally, neither in direct-detection experiments looking for DM scattering off nuclei ~\cite{PandaX-II:2017hlx,PandaX:2018wtu,XENON:2020kmp,XENON:2018voc,LUX-ZEPLIN:2018poe,DARWIN:2016hyl}, via indirect detection via DM annihilation~\cite{HESS:2016mib,MAGIC:2016xys}, nor through direct production in colliders (e.g.~at the LHC~\cite{ATLAS:2017bfj,CMS:2017zts}). This has led to several alternative DM production mechanisms such as the so-called non-thermal production models, in which the observed DM abundance is formed out of equilibrium, either from the annihilation of SM particles via the so-called {\it freeze-in} mechanism~\cite{McDonald:2001vt, Hall:2009bx, Bernal:2017kxu}\footnote{For detection prospects of free-in mechanism, see Refs. \cite{Elor:2021swj,Barman:2022njh,Barman:2021lot}}, inflationary particle production via preheating, direct inflaton decay to DM, or considering the inflaton itself to be DM 
~\cite{Liddle:2006qz,Cardenas:2007xh,Panotopoulos:2007ri,Liddle:2008bm,Bose:2009kc,Lerner:2009xg,DeSantiago:2011qb,Khoze:2013uia,Mukaida:2014kpa,Fairbairn:2014zta,Bastero-Gil:2015lga,Kahlhoefer:2015jma,Tenkanen:2016twd,Daido:2017wwb,Choubey:2017hsq,Daido:2017tbr,Hooper:2018buz,Borah:2018rca,Manso:2018cba,Rosa:2018iff,Almeida:2018oid,Moroi:1994rs, Kawasaki:1995cy, Moroi:1999zb, Jeong:2011sg, Ellis:2015jpg,Harigaya:2014waa, Garcia:2018wtq, Harigaya:2019tzu, Garcia:2020eof,Chung:1998zb, Chung:1998ua, Co:2017mop, Ahmed:2022tfm, Ghoshal:2022jdt}.
 
{Generically, in a non-thermal scenario, DM particles hardly communicate with the visible sector (SM), which makes such scenarios challenging to detect\footnote{Tests of such non-thermal particle production via dark radiation or $N_{\rm eff}$ measurements during the Big Bang Nucleosynthesis (BBN) \& CMB era were proposed in Ref.\cite{Paul:2018njm}. } for any conventional astrophysical or laboratory-based experiment. Thankfully, some of those dark-matter scenarios involve a non-standard evolution of the post-inflationary Universe. In that case, we will argue  that gravitational waves seeded by inflationary tensor perturbations can provide a compelling alternative for probing the existence of such DM production models.}

In our scenario, demanding that the reheaton produces the correct amount of dark matter in the early Universe uniquely dictates the interaction strength of the reheaton with SM and DM particles. The reheaton therefore acts as a portal between the dark and the visible sector. We will therefore study how this portal can lead to sizeable interactions between DM and SM particles, but more interestingly, we will identify regions of the parameter space where the reheaton may be produced in long-lived particle searches experiments such as FASER, MATHUSLA, DUNE, etc.

The paper is organized as follows: In Sec.~\ref{sec:model} we detail the model that will be studied throughout the paper. In Sec.~\ref{sec:GW} we derive the spectrum of GWs that is predicted and compare it to the sensitivity of current and future GW detectors. Considering a minimal model of dark matter interacting with the reheaton, and a higgs portal interaction of the reheaton with the SM Higgs, we explore how the model may lead to smoking-gun signatures in laboratory searches in Sec.~\ref{sec:DM}.

\section{Cosmological Framework}\label{sec:model}
Unlike most of the vanilla inflationary scenario, we consider in this paper the possibility that the inflationary era is followed by a kination-like period with stiff equation of state, which is characterized by an equation of state parameter
\be
w_{\rm kin}>1/3\,.
\ee
Such a cosmological era is typically present in models of non-oscillatory scalar field inflation~\cite{Heurtier:2022rhf, Campos:2004xw, Ellis:2020krl,Feng:2002nb} which feature $w_{\rm kin}\approx 1$. 

The manner in which the Universe is reheated, leading to the subsequent hot big bang era, strongly depends on the model considered. In the context of non-oscillatory scalar-field inflation, it is believed that the SM is reheated through a gravitational production of particles at the end of inflation~\cite{Ford:1986sy,Chun:2009yu,Dimopoulos:2017zvq}. Since the inflationary sector may not be composed of only one scalar field, it may as well be possible that the inflaton transfers a fraction of its energy density into a metastable spectator field $S$, that we call the {\em reheaton}. After inflation, the small oscillations of the reheaton may behave like an extra component of matter in the early universe (corresponding to an equation of state parameter $w_S=0$), which will eventually decay to reheat the SM at a later time. However, if the reheaton potential is different than a quadratic potential, its equation of state parameter $w_S$ could be different than zero. We will study the effect of this parameter on the GW detection, but will restrict our study in the last part of the paper to the standard value $w_S=0$. We will denote by
\be
\eta\equiv \frac{\rho_S}{\rho_{\rm inf}}\ll 1\,,
\ee
the fraction of energy density which is transferred into the reheaton field at the end of inflation. We will also denote by $m_S$ and $\Gamma_S$ the mass and total decay width of the reheaton field. When $H\sim \Gamma_S$ we assume that the reheaton produces both SM and DM particles. We will consider for simplicity that the DM particles produced from this decay, which are initially boosted \mbox{($E_{\rm DM}\approx m_S/2\gg m_{\rm DM}$)}, redshift before composing the cold relic density that is observed today in cosmological data~\cite{Planck:2018vyg}. Demanding that this production accounts for the correct relic abundance fixes the value of the decay branching fraction of the reheaton into DM particles:
\be\label{eq:BrDM}
{\rm Br}\left(S\to {\rm DM},{\rm DM}\right) =\frac{m_S}{2m_{\rm DM}}\left(\frac{\rho_{\rm eq}}{\rho_S}\right)^{1/4}\,,
\ee
where $\rho_S$ denotes the energy density of reheatons at the time of reheating and $\rho_{\rm eq}$ the energy density of the Universe at matter-radiation equality.

The energy density at the end of inflation, the energy fraction released in the form of reheatons at the end of inflation, the decay width of the reheaton, the mass of the reheaton, and the mass of dark matter 
\be
\left\{\rho_{\rm inf}\,,\ w_{\rm kin}\,,\ w_{\rm S}\,,\ \eta\,,\ \Gamma_S\,,\ m_S\,,\ m_{\rm DM}\right\}\,,
\ee
therefore constitute the set of free parameters of the model. 

In Fig.~\ref{fig:drawing} we depict a typical example where the period of kination-like ends before the reheaton has decayed, leading to a period of early matter domination. In order to obtain such an evolution, we simply assume that the different energy components decrease like $a^{-3(w+1)}$. When dark matter is produced through the decay of the reheaton, one should note that it behaves first as radiation, since it is produced with a typical energy $E_{\rm DM}\sim m_S/2$ (where we assume for simplicity a two body decay).

\section{Gravitational-Wave Signatures}\label{sec:GW}

We consider in this section the  first-order tensor perturbations propagating in the early Universe as gravitational waves, created by the inflaton quantum fluctuations ~\cite{Starobinsky:1985ww}. We will study how a modification of the standard cosmological history can affect the spectrum of such GWs and lead to measurable signatures for future gravitational wave detectors.
\subsection{Gravitational-Wave Spectral Shapes from Non-Standard Cosmology}


In standard inflationary scenario, the GW spectrum produced from the inflaton's quantum fluctuations is nearly scale invariant when it is produced and stays that way across the Universe's history as long as the Universe transitions instantaneously from inflation to radiation domination 
\cite{Watanabe:2006qe, Saikawa:2018rcs, Bernal:2019lpc}. 
Nevertheless, given the non-standard evolution that we consider in this paper,  the successive kination-like and matter domination eras before that precede the standard radiation domination era induce interesting structures (inverted triangular shape for our scenario) in the otherwise flat gravitational wave spectrum, hence providing a unique signal of such a non-standard cosmological history.

The energy density stored in the GW spectrum per unit $\log$ interval of $k$ is given by $\frac{d\rho_{GW}}{d\log(k)}=\frac{k^2 h_k^2}{16\pi G a^2}$ \cite{Caprini:2018mtu}, where $h_k$ is the amplitude of $k$-th mode of GW in Fourier space. The GW equation of motion dictates that  $h_k\propto\frac{1}{a}$ after the corresponding $k$-mode enters the Hubble sphere, i.e. when $k>aH$. The $k$ mode which enters the Hubble sphere is related to the scale factor $a$ via $k\propto a^{-\frac{3w+1}{2}}$, $w$ being the equation of state parameter of the dominant component of the Universe during that phase. Therefore, during radiation domination ($w={1}/{3}$), the rate at which the modes re-enter the horizon is identical to the rate at which the GW amplitude decreases, hence keeping the spectral shape unaltered. On the other hand, for $w\ne {1}/{3}$, the aforementioned rates differ, hence tilting the GW spectrum. 
In order to understand how the GW spectrum is affected by the cosmological evolution, and thus the equation of state (e.o.s.) parameter of the Universe $w$ at a given time, one can note that $h_k=h_{k_i}\frac{a_k}{a}$ ($h_{k_i}$ being the initial amplitude of GW $k$ mode when it enters the horizon) and replace $a_k$ by $k^{-\frac{2}{1+3w}}$, giving~\cite{Gouttenoire:2021jhk,Caprini:2018mtu,Haque:2021dha}
\be\frac{d\rho_{GW}}{d\log(k)}\propto a_k^{1-3w}\propto k^{-2\frac{1-3w}{1+3w}}\,.\ee 
The expression clearly dictates that for $w>{1}/{3}$ ($w<{1}/{3}$), the GW spectrum grows (decreases) with the frequency.

In our scenario, inflation is followed successively by a kination-like and an early matter dominated epochs, described by $w_{\rm kin}$ and $w_S$ respectively. As mentioned before, for modes $k_\text{kin}^{0} \leq k \leq k_{\rm end}^{0}$ which enter the horizon during the kination-like era -- between the end of inflation and the end of kination (kin), the GW spectrum features a positive slope  as $w=w_{\rm kin}>\frac{1}{3}$ by definition. Similarly for modes $k_\text{RH}^{0} \leq k \leq k_{\rm kin}^{0}$ entering the horizon during the matter dominated era -- between the end of kination (kin) and the reheating of the Universe operated by the reheaton decay (RH) -- the slope of the GW spectrum is negative.
The mode $k_i$ corresponding to $i$-th transition is related to the scale factor $a_i$ and Hubble $H_i$ at that epoch via the usual relation $k_{i}^{0} =H_i\left(\frac{a_i}{a_0}\right)$. With the analytical approximation where each mode begins oscillating suddenly after the horizon crossing, the GW spectrum may be approximated by a piecewise function given by,

\begin{widetext}
\begin{equation}
	\Omega^{0}_{\rm GW} (k) = \Omega_{\rm GW}^{\rm 0, flat} \begin{cases}
		 1\,, & k < k_\text{RH}^{0} \\ 
		\left(\frac{k}{k_\text{RH}^{0}}\right)^{\frac{2(3w_S-1)}{1+3w_S}}\,, & k_\text{RH}^{0} \leq k \leq k_{\rm kin}^{0} \\
		\left(\frac{k_\text{kin}^0}{k_\text{RH}^{0}}\right)^{\frac{2(3w_S-1)}{1+3w_S}}\left(\frac{k}{k_\text{kin}^{0}}\right)^{\frac{2(3w_{kin}-1)}{1+3w_{kin}}}\,,& k_\text{kin}^{0} \leq k \leq k_{\rm end}^{0}  \\
		0 &  k_{\rm end}^{0}<k
			\end{cases} \,
	\label{eq:GWspecTot}
\end{equation}
\end{widetext}
 Given a particular set of parameters  \mbox{$\left\{\rho_{\rm inf}\,,\ \eta\,,\ \Gamma_S\,,\ m_S\,,\ m_{\rm DM}\,,w_{\rm kin}\,,w_S\right\}$}, $H_i$ and $a_i$ can be determined by fixing the present scale factor to $a_0=1$, and the scale factor at matter-radiation equality $a_eq$ to its measured value \cite{Planck:2018vyg}. Here $\Omega_{\rm GW}^{\rm 0, flat}$ is given by \cite{Opferkuch:2019zbd},

\begin{equation}
\Omega_{\rm GW}^{\rm 0, flat} = \frac{\Omega_{\gamma}^{0}}{24}  \left(\frac{g_{s,\rm eq}}{g_{s,k}}\right)^{\frac{4}{3}} \left(\frac{g_{k}}{g_{\gamma}^{0}} \right) \frac{2}{\pi^{2}} \frac{H_{\rm end}^{2}}{M_P^{2}}\,,
\end{equation}
where $g_{\gamma}^{0} = 2$ and $g_{k}$ is the d.o.f when the $k$ mode re-entered horizon. $H_{\rm end}$ corresponds to the Hubble parameter at the end of inflation. The frequency $f$ of GW is related to a wave-number $k$ via $f=c\frac{k}{2\pi}$, where $c$ is the speed of light.

The total energy density stored under the form of GW's behaves as a radiation-like component and therefore contributes to the total number of relativistic degrees of freedom $N_{\rm eff}$, leading to an extra contribution which can be quantified as
\begin{equation}
\Delta N_{\rm eff} = \frac{8}{7} \left(\frac{11}{4} \right)^{\frac{4}{3}} \frac{\Omega_{\rm GW}^{0}}{\Omega_{\rm \gamma}^{0} }\,,
\end{equation}
where $\Omega_{\rm GW}^{0}$ is defined as
\begin{equation}
\Omega_{\rm GW}^{0} =  \int \frac{df}{f} \Omega_{\rm GW}^{0}(f) \,.
\end{equation}

\subsection{Experimental Sensitivities}

\begin{figure*}
    \centering
    \includegraphics[width=0.48\linewidth]{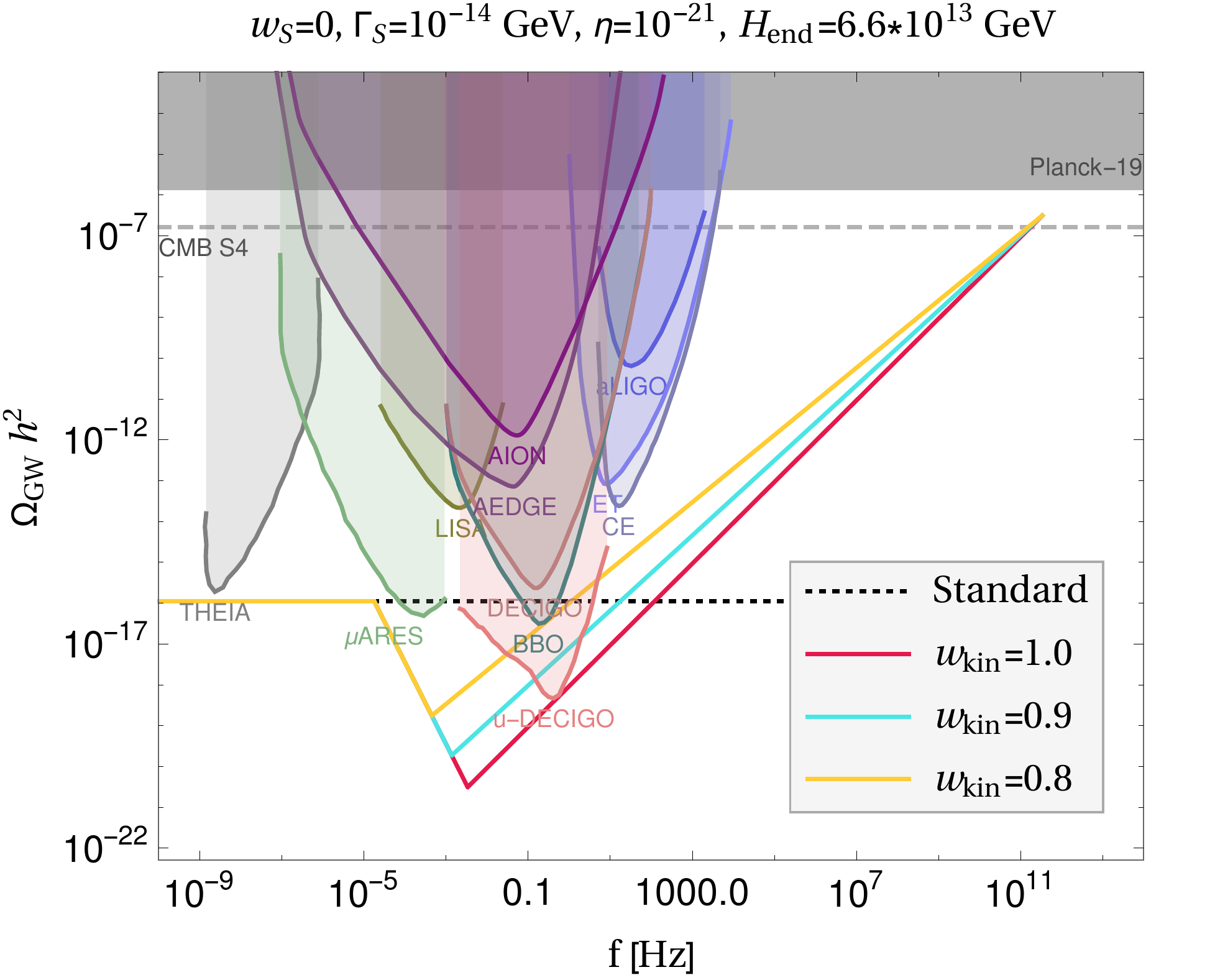}
    \includegraphics[width=0.48\linewidth]{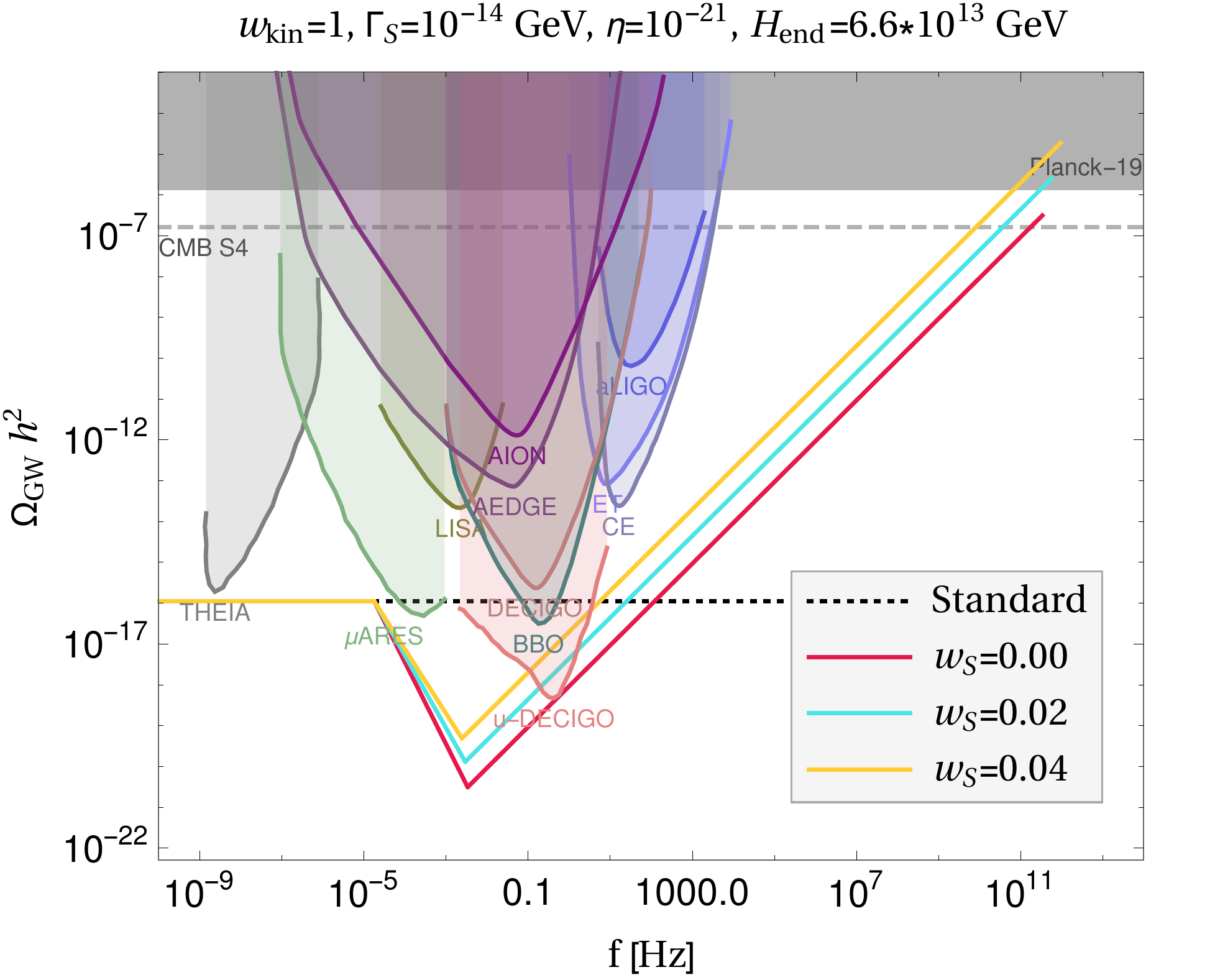}\vspace{0.04\linewidth}
    \includegraphics[width=0.48\linewidth]{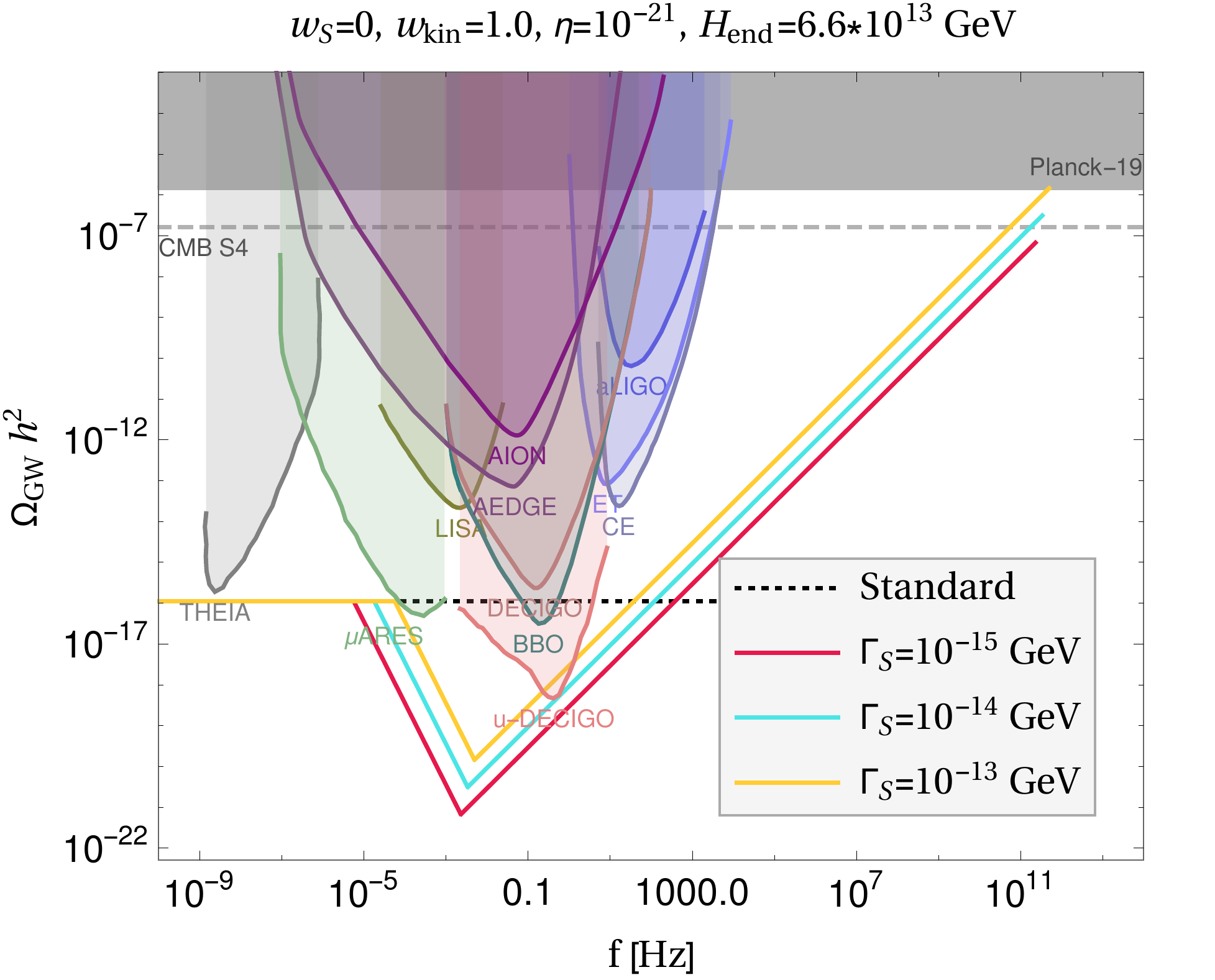}
    \includegraphics[width=0.48\linewidth]{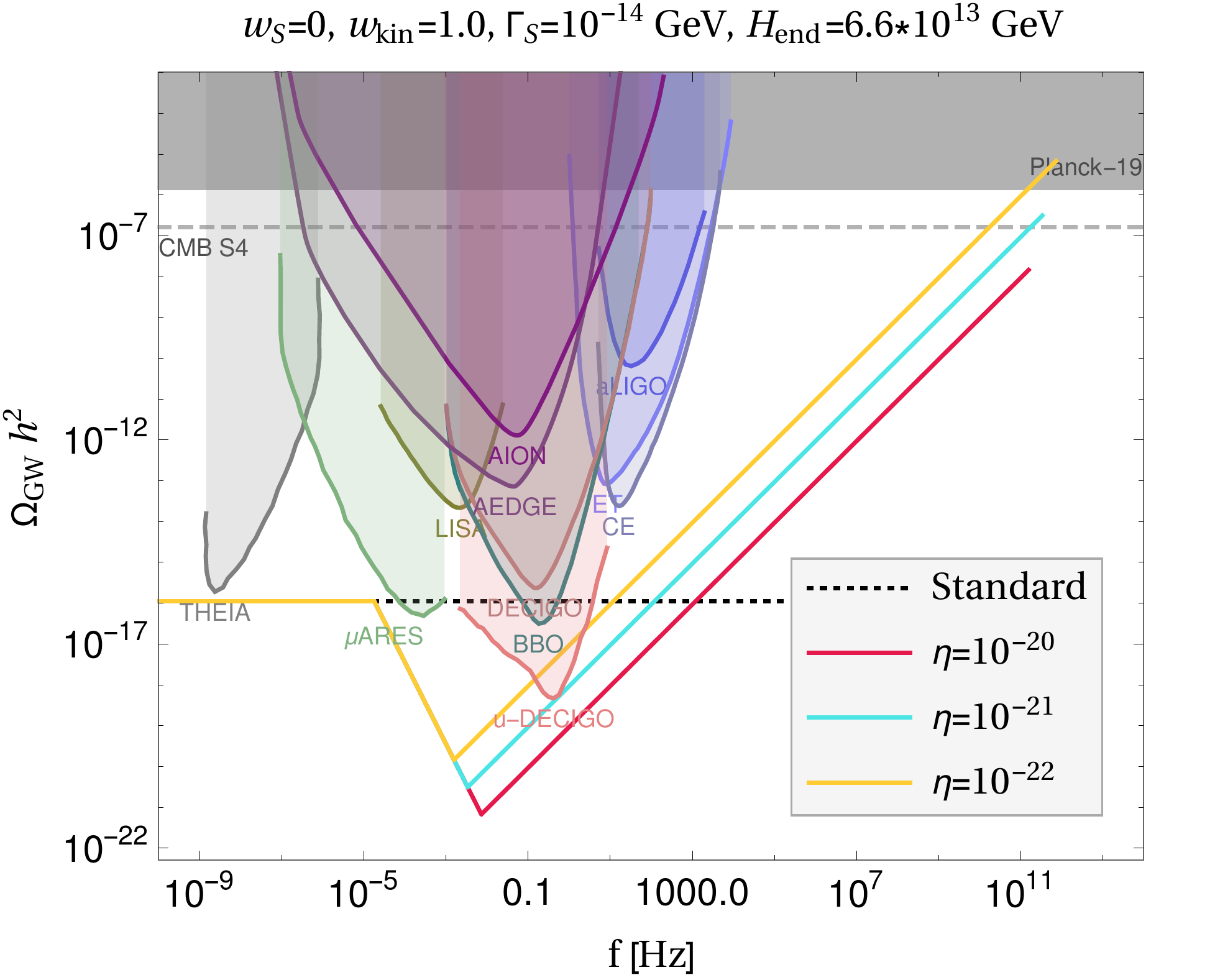}
    \caption{\it \label{fig:GWspec1234} Dependence of GW spectrum with variation of different parameters, $w_k$, $w_S$, $\Gamma_S$, $\eta$ respectively, keeping the other parameters fixed.}
    
\end{figure*}

In fig.\ref{fig:GWspec1234} we show the dependence of the GW spectrum with the different parameters. As we described above, the slope of this spectrum increases during kination domination with higher values of the kination e.o.s parameter $w_{\rm kin}$, with similar effects with the variation of the reheaton e.o.s parameter $w_S$ during the reheaton domination phase. Increasing the decay rate $\Gamma_S$ makes the reheaton decay at an earlier time, resulting in a shorter period of reheaton domination, hence making the ``negative slope region" of the GW spectrum last over fewer e-folds. On the other hand decreasing $\eta$ results in a longer kination dominated period, hence making the ``positive slope region" of GW spectrum larger.

In order to study the sensitivity of present and future GW detectors to our scenario, we have used the sensitivity curves derived in Ref.~\cite{Schmitz:2020syl} for NANOGrav~\cite{McLaughlin:2013ira,NANOGRAV:2018hou,Aggarwal:2018mgp,Brazier:2019mmu}, PPTA~\cite{Manchester:2012za,Shannon:2015ect}, EPTA~\cite{Kramer:2013kea,Lentati:2015qwp,Babak:2015lua}, IPTA~\cite{Hobbs:2009yy,Manchester:2013ndt,Verbiest:2016vem,Hazboun:2018wpv}, SKA~\cite{Carilli:2004nx,Janssen:2014dka,Weltman:2018zrl}, LISA~\cite{LISA:2017pwj, Baker:2019nia}, BBO~\cite{Crowder:2005nr,Corbin:2005ny,Harry:2006fi}, DECIGO~\cite{Seto:2001qf,Kawamura:2006up,Yagi:2011wg}, CE~\cite{LIGOScientific:2016wof,Reitze:2019iox} and ET~\cite{Punturo:2010zz, Hild:2010id,Sathyaprakash:2012jk, Maggiore:2019uih}, $\mu-$ARES~\cite{Sesana:2019vho}, GAIA, THEIA ~\cite{Garcia-Bellido:2021zgu} and aLIGO and aVirgo~\cite{Harry:2010zz,LIGOScientific:2014pky,VIRGO:2014yos,LIGOScientific:2019lzm}.

\begin{figure*}
    \centering
    \includegraphics[width=0.48\linewidth]{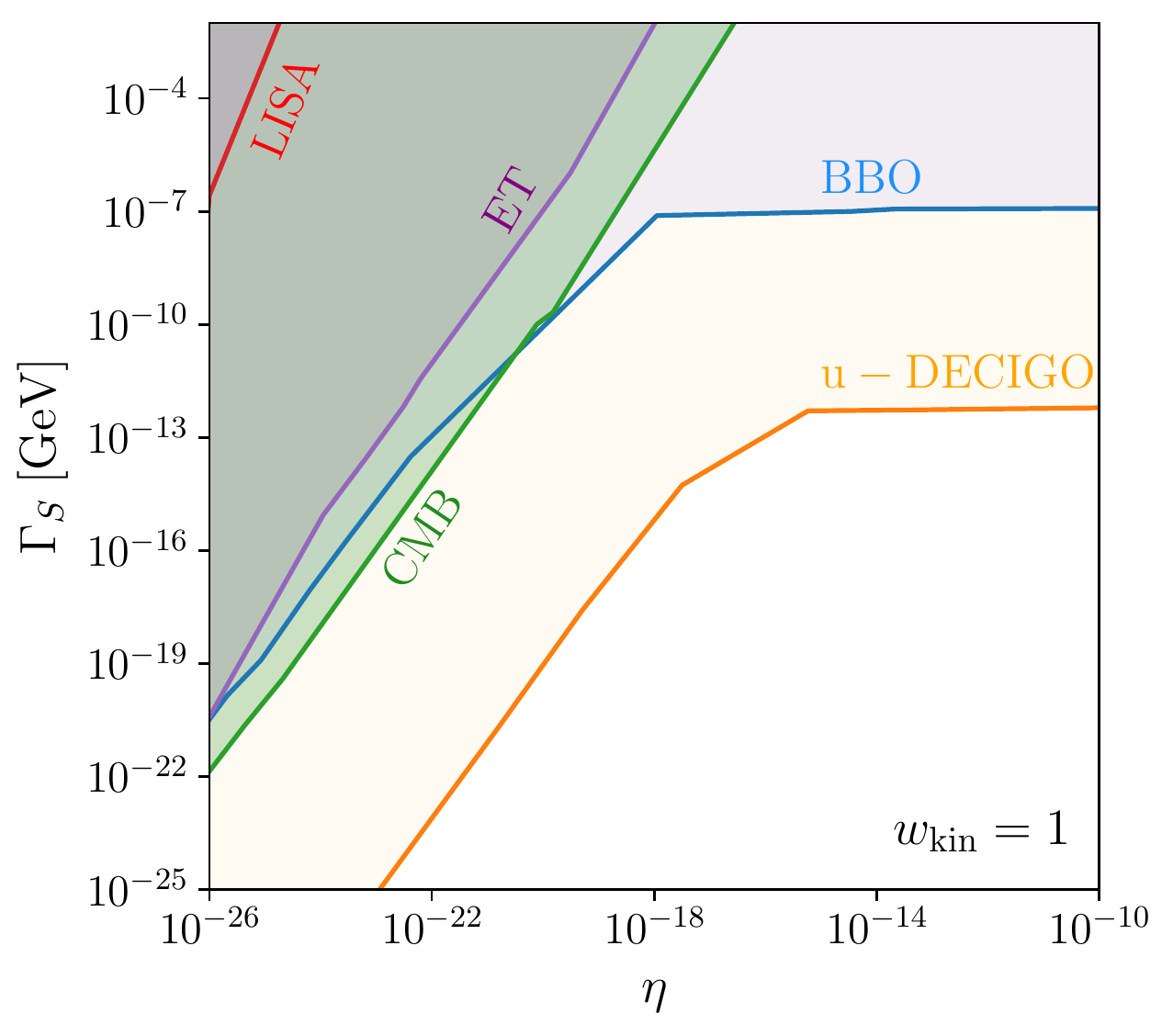}\hspace{0.04\linewidth}\includegraphics[width=0.48\linewidth]{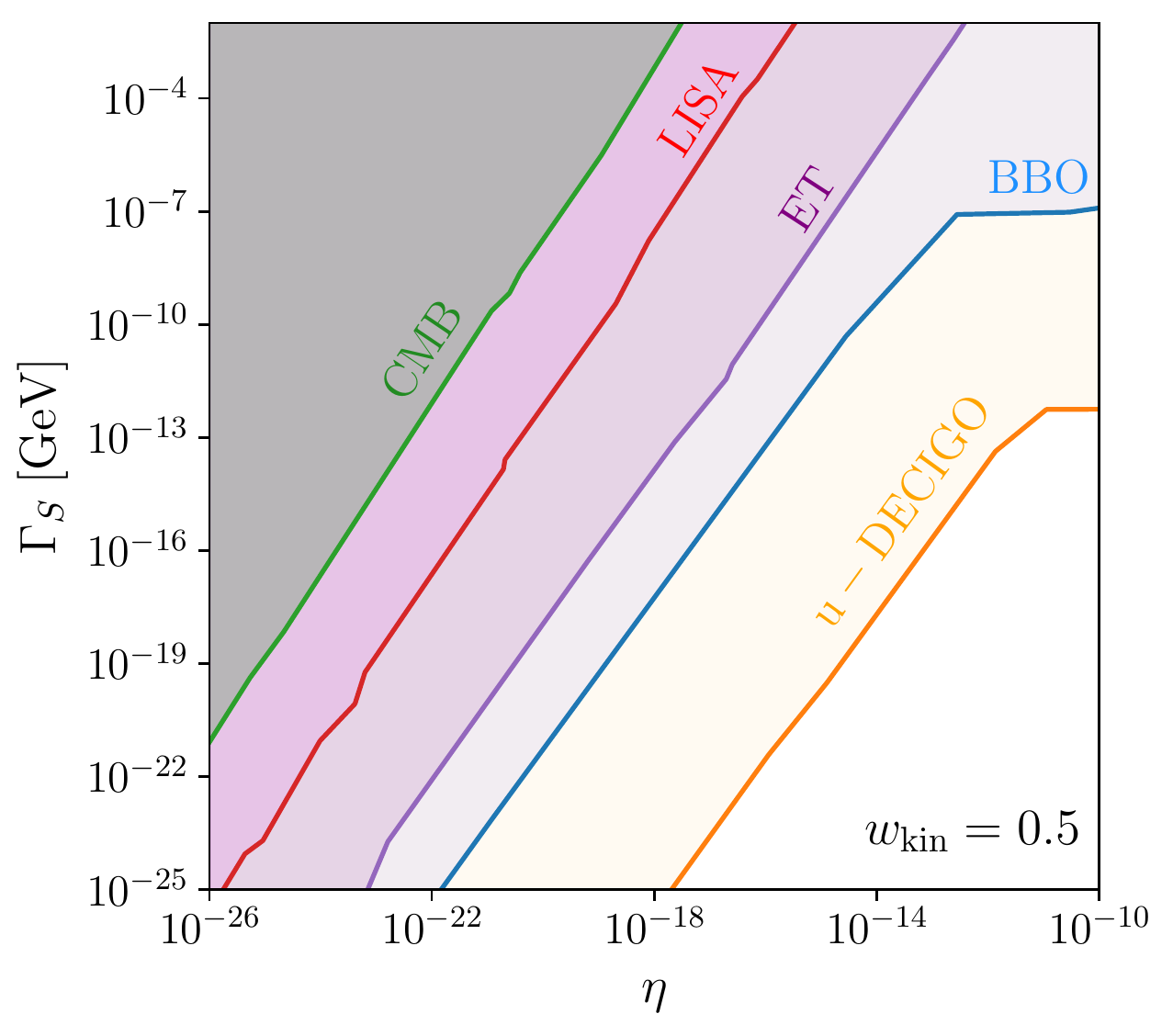}
    \caption{\it \label{fig:GW} The allowed region of $\Gamma_S-\eta$ parameter space from $N_{\rm eff}$ bounds and future reaches in GW detectors (LISA, ET, BBO, u-DECIGO) after 4 years of exposure. The green shaded region above the green line (denoted by CMB) is disallowed by the $N_{\rm eff}<0.3$ constraint. The other colored lines correspond to SNR$=1$ for different GW detectors, the region above then denoting SNR$>1$. The left and right panel correspond to $w_{\rm kin}=1$ and $0.5$ with the other parameters fixed at $\left\{ w_S,~ H_{\rm end},~ m_{\rm DM},~m_S \right\}$ =  $\left\{0,~ 6.6\times10^{13}\rm{GeV}, ~10^8\rm{GeV},~10^{12}\rm{GeV}\right\}$.}
    
\end{figure*}

The signal-to-noise ratio (SNR) ~\cite{Thrane:2013oya,Caprini:2015zlo}
 \begin{align}
 	\text{SNR}_\text{exp} \equiv \left\{ 2 t_\text{obs} \int_{f_\text{min}}^{f_\text{max}} d f \left[ \frac{h^2 \Omega_\text{GW}(f)}{h^2\Omega_\text{eff}(f)}\right]^2\right\}^{1/2}\,.
 \end{align}
 Here $t_\text{obs}$ denotes the observation time and $\Omega_\text{eff}(f)$ corresponds to the noise curve of the GW detector working between the frequency interval $f_\text{min}$ to $f_\text{max}$.
In fig. \ref{fig:GW} we show the visibility (SNR$>1$) range of the GW spectrum. As observed previously in fig.~\ref{fig:GWspec1234}, both increasing $\Gamma_S$ and decreasing $\eta$ brings the GW spectrum more into reach of the GW detectors. This fact is mirrored in fig. \ref{fig:GW} also, as the SNR increases top-left from bottom-right corner. A sudden change of slope of the SNR=1 lines are visible for the GW detectors BBO and u-DECIGO. This means that beyond a certain large $\Gamma_S$, increasing $\eta$ after a certain threshold does not change SNR considerably. This happens when $f_{\rm RH}$ reaches the detector sensitivity curves from the left (with increasing $\Gamma_S$) in \ref{fig:GWspec1234} and $f_{\rm kin}$ is such that the spectrum is substantially below the reach of GW detectors. 






\section{Complementary Probe of Dark Sector}\label{sec:DM}
For a choice of parameters \mbox{$\left\{\rho_{\rm inf}\,,\ \eta\,,\ \Gamma_S\,,\ m_S\,,\ m_{\rm DM}\right\}$}, we have seen in Sec.~\ref{sec:GW} that the peculiar evolution of the post-inflationary Universe may lead to detectable signatures in the gravitational wave spectrum. Furthermore, as we noticed in Sec.~\ref{sec:model}, the decay branching fraction of the reheaton into DM particles is uniquely given by the choice of parameters of the model. For a given microscopic model, knowing the value of the total decay width of the reheaton together with its decay branching fractions into SM and DM particles is equivalent to knowing its interaction strength with either species. In this section, we therefore introduce a specific particle physics model for the dark sector and investigate whether dark-particle searches may provide us with smoking-gun signatures that are complementary to gravitational wave searches.

\subsection{Microscopic Model: higgs-portal DM}

As the simplest extension of the SM that contains a dark matter particle and an additional singlet scalar field, we consider a higgs-portal scenario in which the dark-matter particle is a Dirac fermion $\chi$, and couples to the reheaton $S$ through a Yukawa coupling. We denote by $g_\chi$ the coupling of such interaction and consider that the reheaton mixes with the SM Higgs boson $H$ with mixing angle $\sin\theta$. Before rotation into the mass eigenstate basis, the lagrangian is of the form
\be
\mathcal L \supset -g_\chi S\bar \chi\chi -V_{\rm SM}(H)- V_{\rm reh}(S) - \lambda_{HS}|H|^2|S|^2\, + H.c..
\ee
After rotation into the mass-eigenstate basis, and in the limit of small mixing angle, the mass eigenstates can be simply approximated to be
\bea
\tilde H &\approx& H - S\sin\theta\,, \nonumber\\
\tilde S &\approx& S + H\sin\theta \,.
\eea
where 
\be
\tan 2\theta =  \frac{2 v_S v_h\lambda_{HS}}{(m_H ^2 - m_S ^2)}
\ee
$v_S$ being the vev of the S field and $v_S$ is the EW vev. In the scenario that we have studied previously, we have considered models in which the reheaton is long-lived, may dominate the energy density of the Universe, and eventually decays into SM and DM particles. Demanding that this decay is kinematically allowed implies that  the dark-matter particle must be lighter than the reheaton
\be
m_{\rm DM}<\frac{m_S}{2}\,.
\ee
Furthermore, the reheaton is long-lived, and is the only mediator between the visible and the dark sector in this scenario. This, in turn, means that DM is very feebly coupled to the SM bath and does not thermilize throughout the history of the Universe. When presenting our results, we shall verify this assumption, since we have demanded that the reheaton produces out-of-equilibrium the whole relic density of dark matter in the previous sections. Because of its small cross section of annihilation into and scattering with SM particles, it is expected that the direct or indirect detection of DM is beyond the reach of laboratory experiments and astrophysical searches. However, the reheaton, which is the mediator between the dark and the visible sector, may couple to SM particles more strongly and may show some signatures in long-lived particle (LLP) searches. In particular, long-lived particles with masses $\lesssim 5 \mathrm{GeV}$ may be detectable soon with experiments such as FASER and FASER-II \cite{Feng:2017vli,FASER:2018eoc,FASER:2018bac,FASER:2019aik} , DUNE \cite{DUNE:2015lol,Berryman:2019dme}, DarkQuest-Phase 2 \cite{Batell:2020vqn}, MATHUSLA \cite{Curtin:2018mvb}, PS191 \cite{Bernardi:1985ny,Gorbunov:2021ccu} or SHIP \cite{SHiP:2015vad}. 
In this range of masses, for each point in the parameter space, demanding that the reheaton produces the correct relic density of DM provides via Eq.~\ref{eq:BrDM} the decay branching fraction of the reheaton into dark matter. Given the value $\Gamma_S$, this provides us with the value of the decay width of the reheaton into DM and SM particles. Because we consider a reheaton which is lighter than the SM Higgs boson, it is easy to obtain from these decay widths what the mixing angle $\sin\theta$ is by simply writing
\be
\Gamma(S\to SM) \equiv \sin\theta^2 \Gamma_H(m_S)\,,
\ee
where $\Gamma_H(m_S)$ simply denotes the decay width of the SM Higgs boson restricted to channels which are kinematically open for the reheaton and rescaled by a factor $m_S/m_H$. Details on microscopic models of the Higgs portal for a fermionic dark matter can be found in Refs.~\cite{Okada:2019opp, Arcadi:2021mag}.
\subsection{Dark-Matter Non-Thermal Production}
Before we present our results, a few comments are in order. First, as we just mentioned, we have assumed throughout this paper that the DM particle is only produced out of equilibrium, through the decay of the reheaton. While scanning over the parameter space, it is possible that one reaches a regime in which the interactions of DM with the SM may lead to the thermalization of dark matter, or to a sizeable production of DM particles out of equilibrium through thermal processes a la Freeze-In. In order to avoid these possibilities, we used a modified version of the code developed by the authors of Refs.~\cite{Barman:2022njh,Barman:2021lot} to estimate the amount of dark matter produced out of equilibrium via two-to-two processes. When the reheaton decays and reheats the Universe, it may as well be possible that the reheating temperature exceeds the reheaton mass and lead him to be close from thermalizing. In order to ensure that this is not the case, we excluded from our scans points with a reheating temperature larger than the reheaton mass. Finally, we also excluded, of course, points for which the reheating temperature is below the temperature of BBN, dark-matter masses below $5$keV (to avoid constraints on warm dark matter such as Lyman-$\alpha$ constraints or the Tremaine-Gunn bound), and points for which DM is relativistic during BBN and is excluded by observational constraints on $\Delta N_{\rm eff}$.
\begin{figure*}
    \centering
    \includegraphics[width=0.48\linewidth]{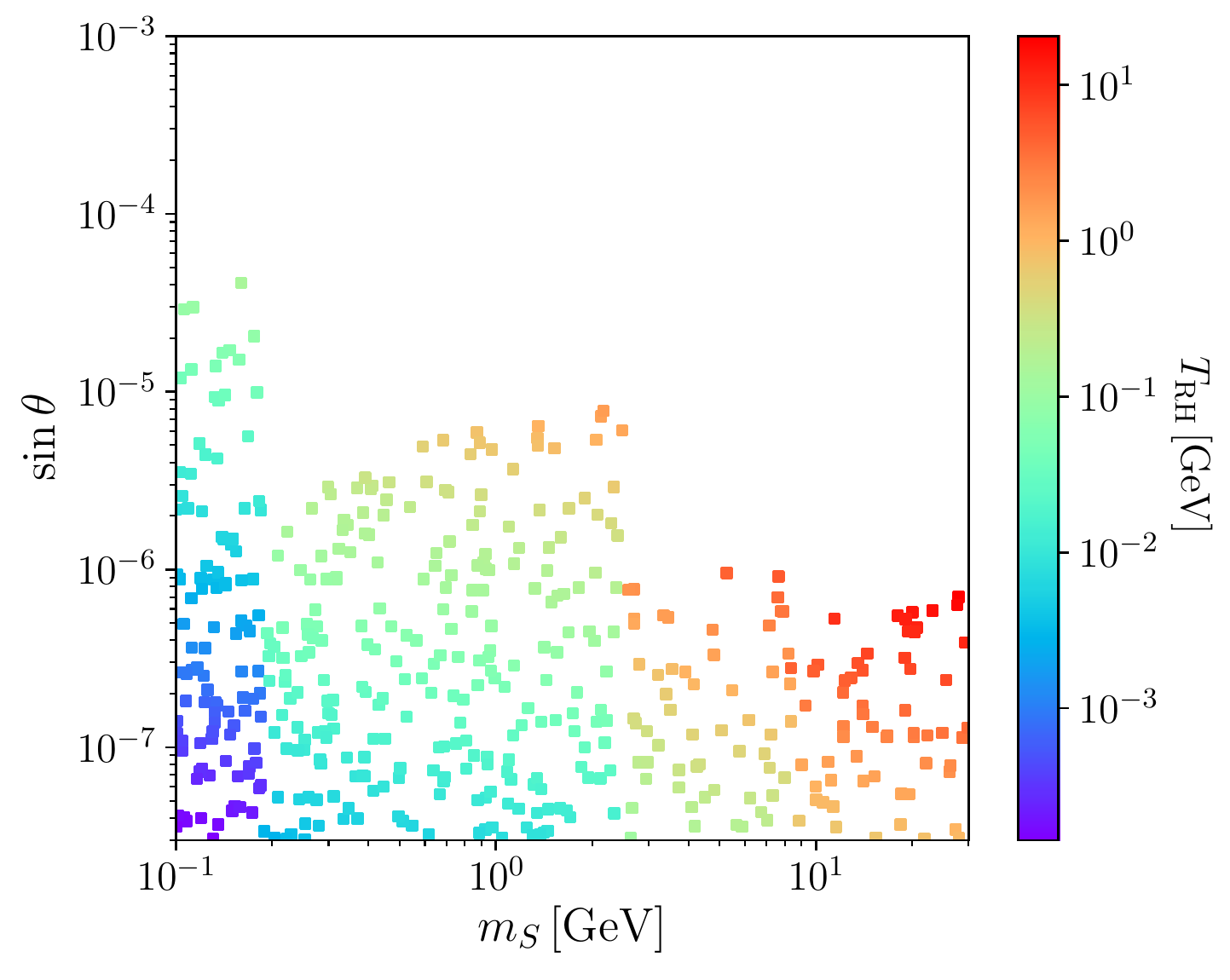}\hspace{0.04\linewidth}\includegraphics[width=0.48\linewidth]{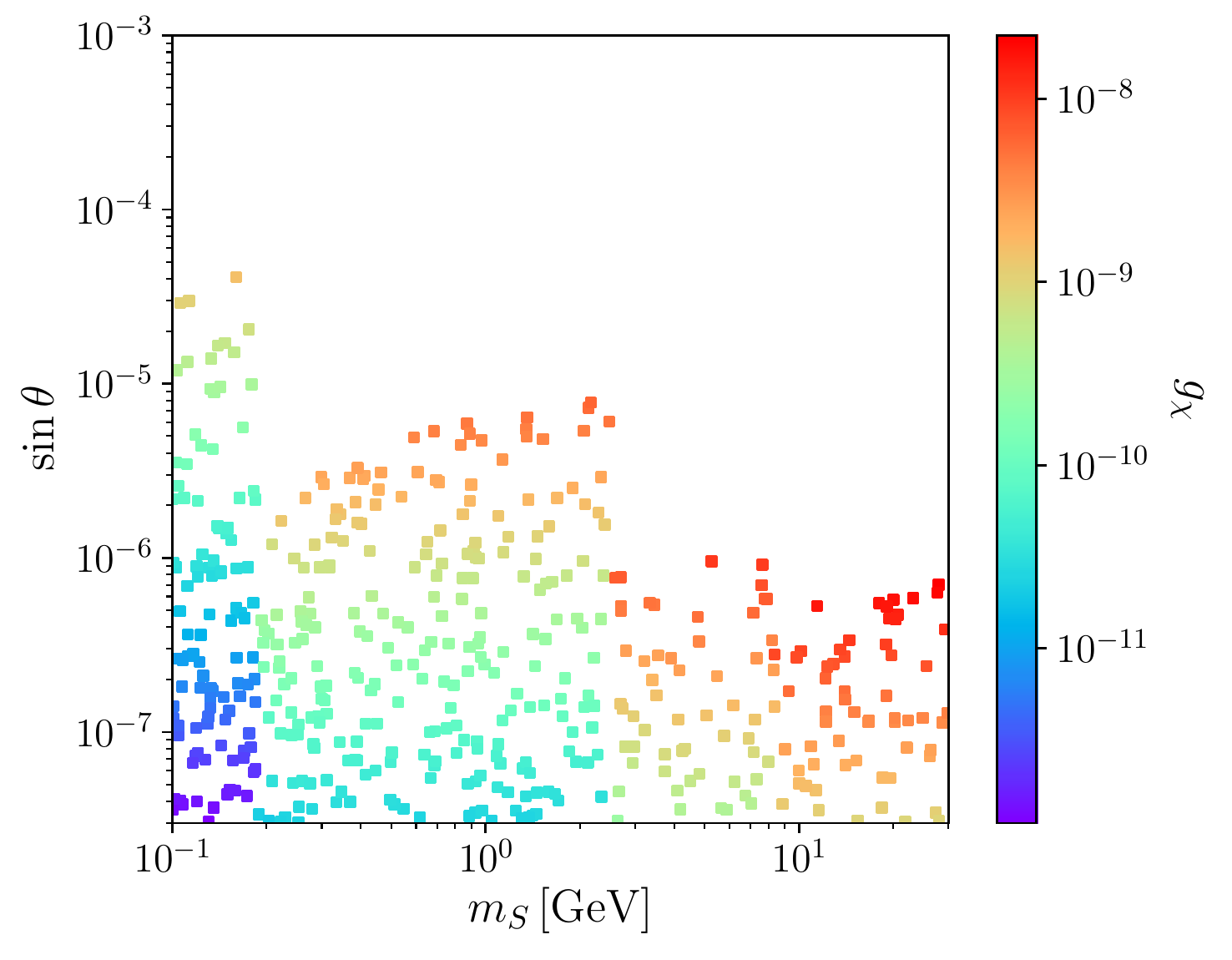}
    \caption{\it \label{fig:DM_scan} Scan over the energy fraction $\eta$, the reheaton decay width $\Gamma_S$ and the dark-matter mass $m_{\rm DM}$. We assumed $w_{\rm kin}=1$ and excluded points for which the Freeze-In production of DM exceeds 1\% of the total relic density.
     }
\end{figure*}

In Fig.~\ref{fig:DM_scan} we present our results in the plane $(m_S,\sin\theta)$ and exhibit the value of the reheating temperature (left panel) and the dark sector coupling g$_{\chi}$ (right panel) for all points surviving the various constraints mentioned above, and in the case $w_{\rm kin}=1$. As one may see, points with a large mixing angles are excluded since they would lead to a sizeable production of dark matter from thermal processes which could either overclose the Universe or lead dark matter to thermalize. Interestingly, although the value of the dark coupling is quite small, the value of the mixing angle between the reheaton and the SM Higgs can remain sizeable to make detection prospects good.  

The variation of the reheating temperature and the dark coupling on those plots can be understood as follows: The larger the mass of the reheaton, the larger is its decay width and therefore the larger is the reheating temperature. Moreover, for a fixed cosmological evolution, hence a fixed value of $\eta$, $\Gamma_S$ and $m_S$, a larger $\sin \theta$ leads to a smaller branching fraction of the reheaton decay into DM, which has to be compensated by a larger dark coupling. 

\medskip

\section{Results}

\begin{figure*}
    \centering
    \includegraphics[width=0.47\linewidth]{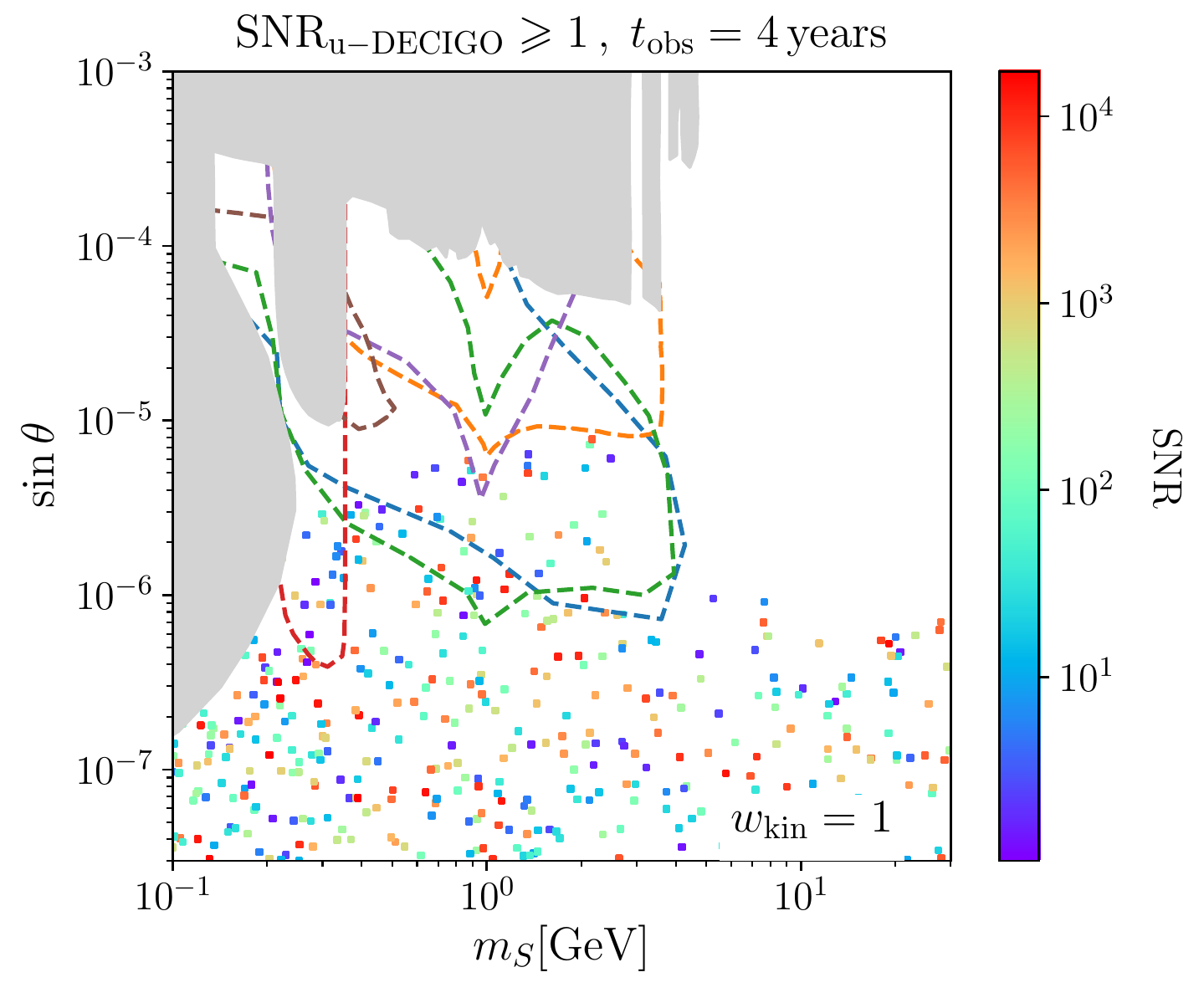}\hspace{0.04\linewidth}\includegraphics[width=0.49\linewidth]{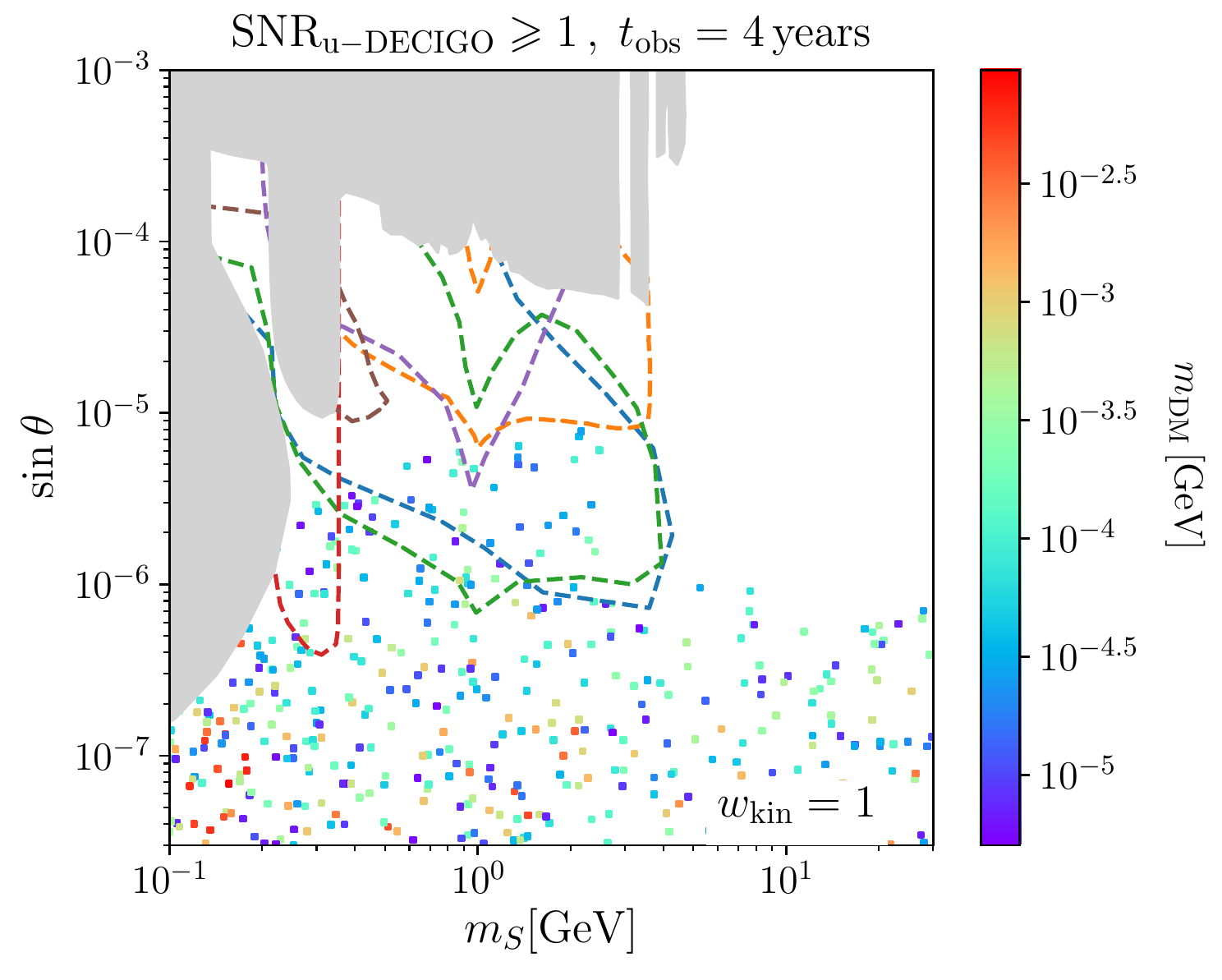}
    \caption{\it \label{fig:complementary} \footnotesize Comparison between the points in parameter space which will detected by u-DECIGO after 4 years of exposure (with SNR$\geqslant 1$) and the sensitivity limits which will be probed in future long-lived particle searches. The dashed contours depict such regions for FASER II (orange)  \cite{Feng:2017vli,FASER:2018eoc,FASER:2018bac,FASER:2019aik} , DUNE (red) \cite{DUNE:2015lol,Berryman:2019dme}, DarkQuest-Phase 2 (purple) \cite{Batell:2020vqn}, MATHSULA (green) \cite{Curtin:2018mvb}, PS191 (brown) \cite{Bernardi:1985ny,Gorbunov:2021ccu}, and SHIP (blue) \cite{SHiP:2015vad}. We considered the case $w_{\rm kin}=1$ in both plots and the color bar indicates the values of the DM mass in the left panel, and the value of the expected SNR in the right panel\footnote{The region on the right of the shaded grey for masses 5 GeV and above although shown in white but there maybe some future experiments there, however we do not consider them since our parameter space does not contain points in those regions, for more see \cite{Das:2022oyx}.}.
    }
    
\end{figure*}

\begin{figure*}
    \centering
    \includegraphics[width=0.47\linewidth]{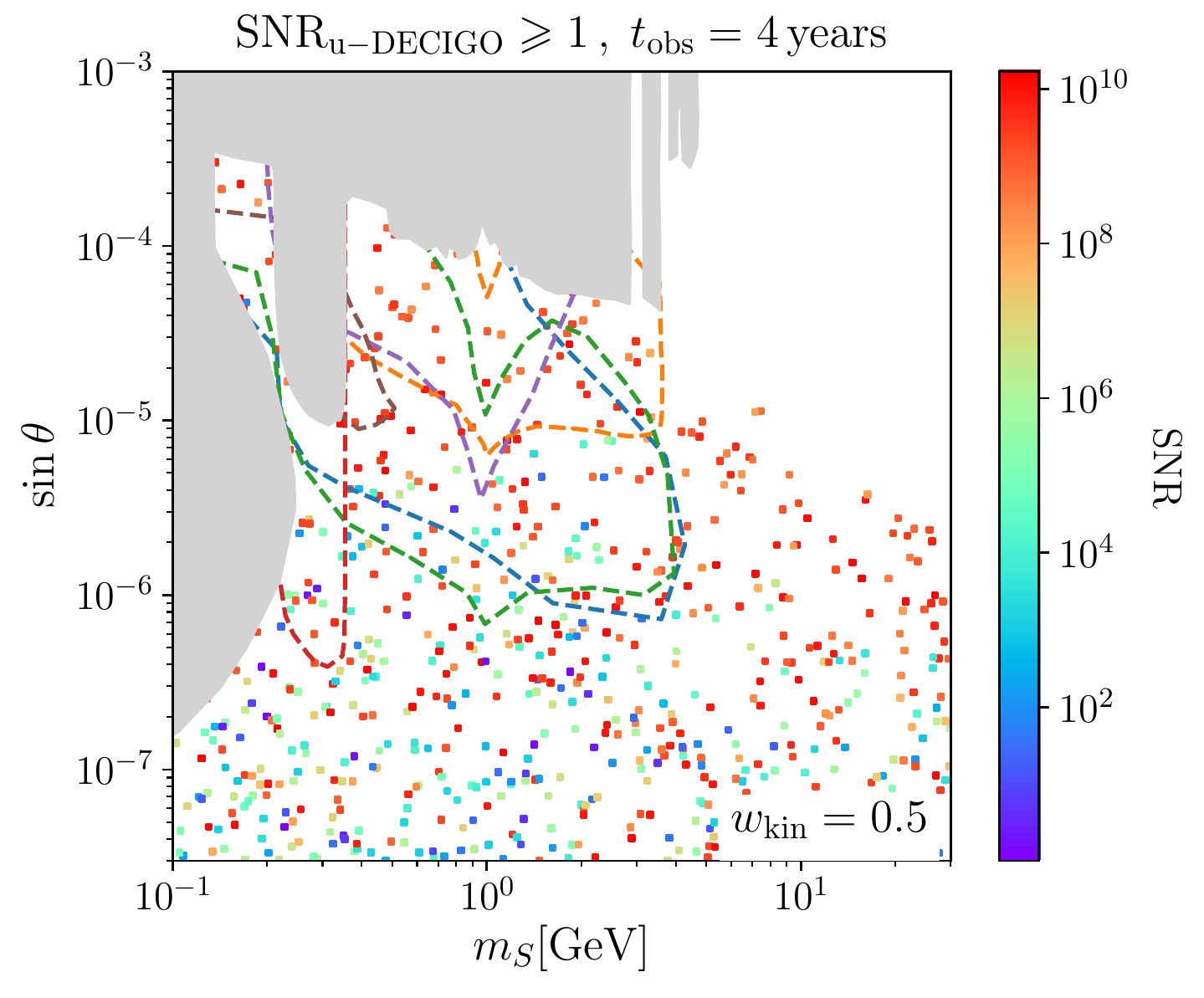}\hspace{0.04\linewidth}\includegraphics[width=0.49\linewidth]{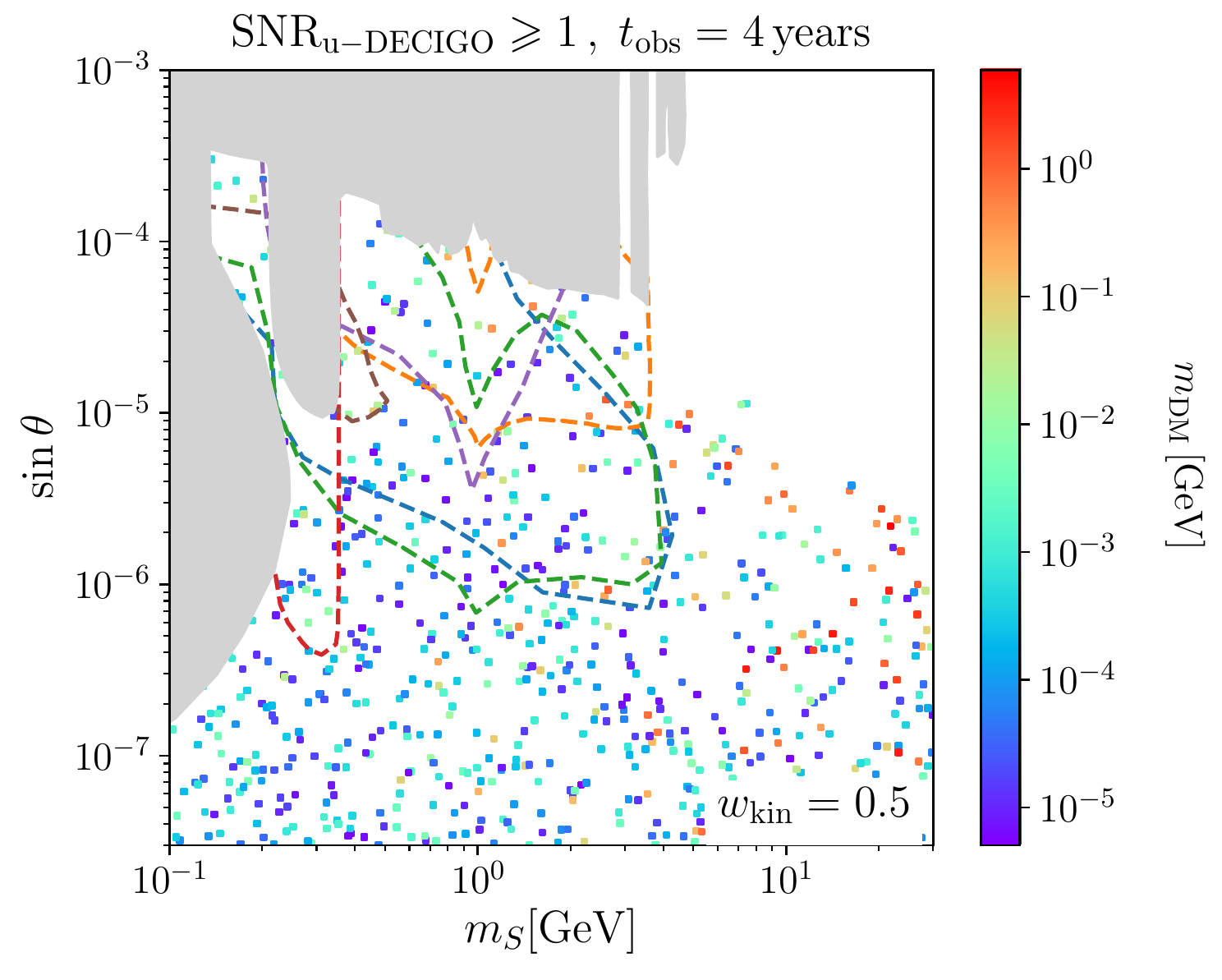}
    \caption{\it \label{fig:complementary05} \footnotesize Same as Fig.~\ref{fig:complementary} but for $w_{\rm kin}=0.5$.
    }
    
\end{figure*}

\begin{figure*}
    \centering
    \includegraphics[width=0.47\linewidth]{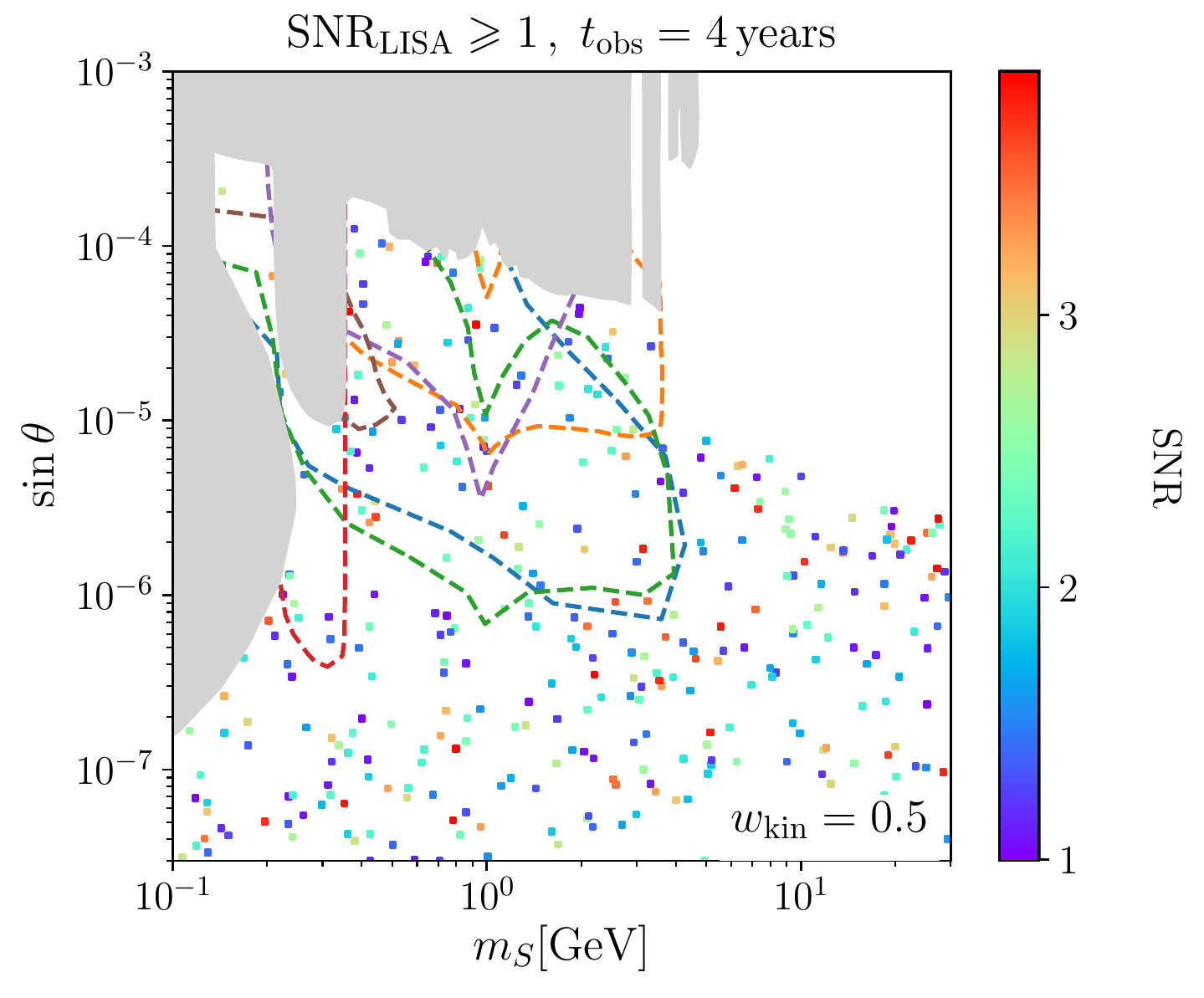}\hspace{0.04\linewidth}\includegraphics[width=0.49\linewidth]{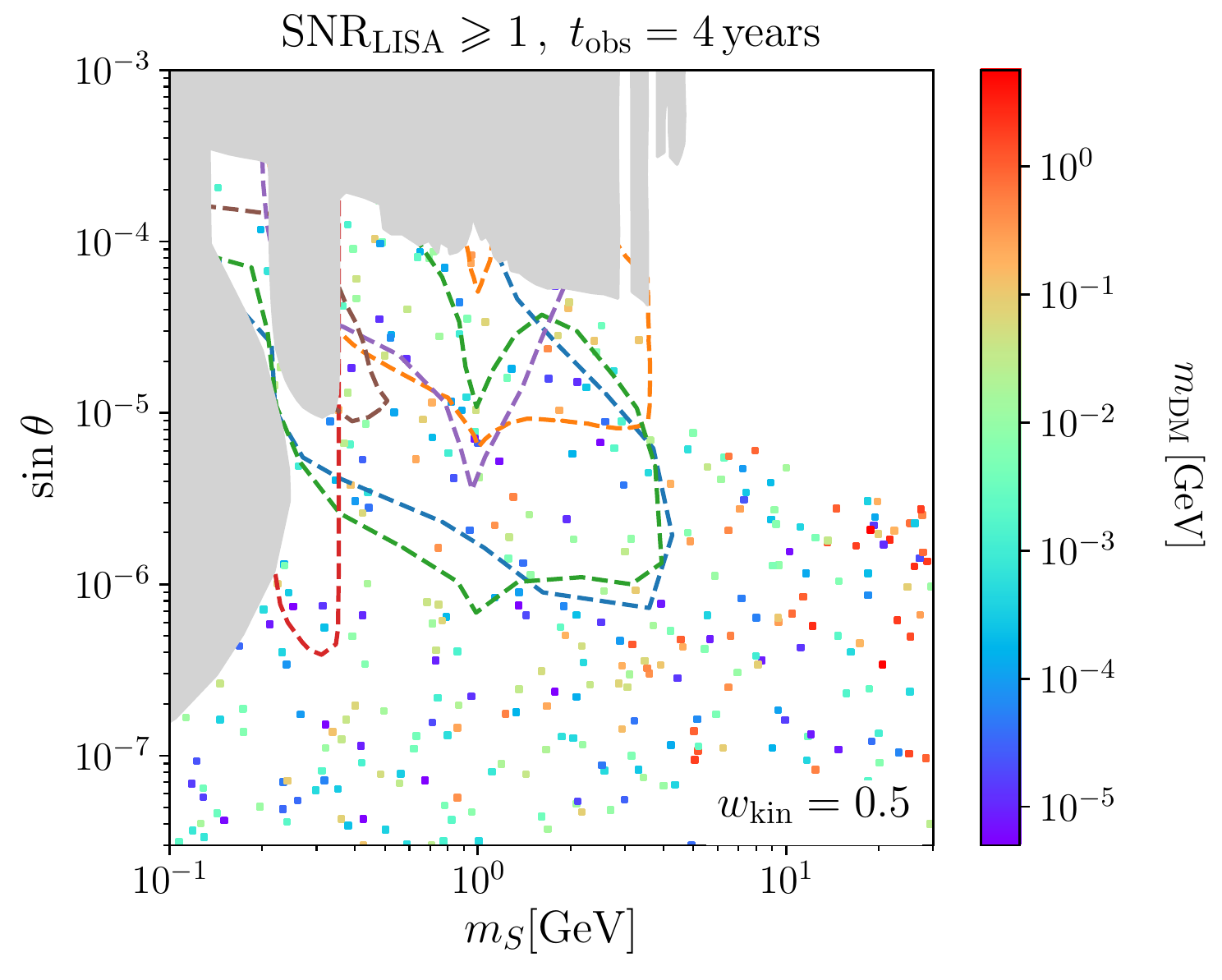}
    \caption{\it \label{fig:complementary05LISA} \footnotesize Same as Fig.~\ref{fig:complementary} but for $w_{\rm kin}=0.5$ and an SNR computed for LISA.
    }
    
\end{figure*}
As we could see in Sec.~\ref{sec:GW}, a large fraction of the parameter space of our model may be probed by future GW detectors in the future. For every point in the corresponding parameter space, we have seen in Sec.~\ref{sec:DM} that it is straightforward to derive, for a given dark matter model, the value of the reheaton coupling with SM particles in order to obtain the correct relic density of DM particles in the present Universe. In Fig.~\ref{fig:DM_scan} it appears that the value of this coupling can be sizeable, which makes laboratory experiments compelling in order to search for the existence of such a reheaton particle.

In Fig.~\ref{fig:complementary}  we restrict the scan presented in Fig.~\ref{fig:DM_scan} to points which would lead to an SNR larger than one for u-DECIGO after 4 years of operation. We then compare our results to the current limits on long-lived particles interacting with the SM and to the future limits which may be reported in the near future by collaborations such as FASER \cite{Feng:2017vli,FASER:2018eoc,FASER:2018bac,FASER:2019aik} , DUNE \cite{DUNE:2015lol,Berryman:2019dme}, DARKQUEST-2 \cite{Batell:2020vqn}, MATHUSLA \cite{Curtin:2018mvb}, PS191 \cite{Bernardi:1985ny,Gorbunov:2021ccu}, and SHIP \cite{SHiP:2015vad}. In Fig.~\ref{fig:complementary05} and \ref{fig:complementary05LISA}, similar results are presented for $w_{\rm kin}=0.5$ for points which feature an SNR larger than one for u-DECIGO and LISA, respectively.

Note that in the case of $w_{\rm kin}=1$, however, for BBO, or detectors with a lower sensitivity,  the situation is different. Indeed, as one can see from the left panel of Fig.~\ref{fig:GW}, for the case of $w_{\rm kin}=1$ that we considered the region of parameter space which can be probed by BBO within 4 years of exposure and which is not excluded by CMB measurement lies at relatively large values of the reheaton's total decay width. Therefore, points which may be visible by BBO appear to lead to a too large production of DM particles through Freeze-In or even the thermalization of DM.
\medskip

\section{Discussion and Conclusion}

Various sources such as first-order phase transitions, cosmic strings or domain walls, inflationary preheating, etc. are predicted to produce detectable gravitational waves from early Universe. The detection of such signals will opens up a compelling new window into the pre-BBN Universe. In fact it can even help probing new physics beyond the SM, as for example GUT-scale physics, high scale baryogenesis and leptogenesis physics \cite{Dasgupta:2022isg,Bhaumik:2022pil,Barman:2022yos,Ghoshal:2022jdt,Dunsky:2021tih,Bernal:2020ywq,Ghoshal:2020vud} which are otherwise beyond the reach of LHC or other laboratory or astrophysical searches for new physics. In this work, we studied the non-thermal production of dark matter within a non-standard cosmological framework with large GW signals arising from first-order inflationary tensor perturbations, the GW spectrum of which are different from each other cosmic sources described above\footnote{For non-thermal DM search with GW from preheating, see \cite{Ghoshal:2022jdt}}.  
We considered the simple case where, after the end of inflation, a fraction of energy $\eta$ is transferred to the oscillation of a scalar field (for example, a moduli field) which we dub as the reheaton $S$, whereas the Universe undergoes a kination or kinetion-like phase with $w_{\rm kin}>1/3$. We investigated the case where dark matter is simply produced from the decay of the reheaton, which also reheats the Universe by decaying into SM particles. The reheaton may decay during kination domination, or after, leading to a phase of early matter domination after the kination era. The entire setup is therefore described by a minimal set of independent parameters: the mass and decay width of the reheaton $m_S$ and $\Gamma_S$, the energy fraction of reheaton particles at the end of inflation $\eta$, the equation of state parameters of the kination and reheaton domination phases $w_{\rm kin}$ and $w_S$, and the dark-matter mass $m_{\rm DM}$. Assuming a maximal amplitude of scale-invariant tensor modes produced during inflation, as allowed by the current CMB measurements, we studied the effect of those different parameters on the shape of the gravitational wave spectrum at present time. We showed that such a non-standard cosmological history leaves imprints in GW signals, to be measured by upcoming detectors. We exhibited the GW spectral shapes that may be observed in LISA, ET, BBO, DECIGO and u-DECIGO.  We showed in particular that models with $ w_S=0$ (which corresponds to a reheaton oscillating in a quadratic potential) would show detectable signals for BBO and u-DECIGO for $w_{\rm kin}=1$, but may also be probed by LISA and the ET for lower $w_S\lesssim 0.5$ as can be seen from Fig.~\ref{fig:GW}.

Because the production of the correct DM relic density is intrinsic to our model and implicitly given by the lifetime and decay branching fraction of the reheaton into DM particles, the interaction strength of the reheaton with DM and SM particles can be simply extracted from each point in the parameter space. We showed that the reheaton can therefore be searched for experimentally in light dark sector searches involving intensity, lifetime and beam dump experiments. We explored the parameter space in Fig.~\ref{fig:complementary}-\ref{fig:complementary05LISA} and compared our results to the projected limits of laboratory searches such as FASER \cite{Feng:2017vli,FASER:2018eoc,FASER:2018bac,FASER:2019aik} , DUNE \cite{DUNE:2015lol,Berryman:2019dme}, DARKQUEST-2 \cite{Batell:2020vqn}, MATHUSLA \cite{Curtin:2018mvb}, PS191 \cite{Bernardi:1985ny,Gorbunov:2021ccu}, and SHIP \cite{SHiP:2015vad}. In particular, we found that a kination-like period with equation-of-state parameter $w_{\rm kin}\approx 0.5$, a reheaton mass $\mathcal O(0.5-5)$ GeV, and a DM mass of $\mathcal O (10-100)$ keV could show detectable signals for DUNE, MATHUSLA, and SHIP. Interestingly, we also showed that this same region of parameter would show detectable signals within 4 years of exposure for GW detectors such as LISA and u-DECIGO, as a complementary smoking-gun signature.

When studying the testability of our non-thermal DM production mechanism, we used a simple Higgs-Portal set-up and a fermionic DM particle. However, we insist on the fact that our prescription to search for complementary probes of new physics with laboratory and Gravitational Wave experiments is very general and can be applied to many other DM scenarios that involve a non-standard cosmology \cite{Ghoshal:2021ief,Barman:2022njh,Banerjee:2022fiw,DEramo:2017gpl,Co:2015pka,Heurtier:2019eou, Heurtier:2017nwl,Heurtier:2022rhf}. Following this prescription, we believe that many realizations of non-thermal DM production in the early universe may lead to unique predictions of GW spectral shapes that can be detected in future GW experiments and searched for experimentally in laboratories.


It is remarkable that the existence of a non-standard post-inflationary cosmology plays a crucial role in shaping the morphology of the gravitational wave spectrum for a given microscopic particle physics scenario. Importantly, this post-inflationary story-line may also leave imprints in the CMB spectrum itself. Indeed, it affects the number of $e$-folds of inflation, and may lead to refine predictions for the inflation observables strongly correlated to astrophysical signals that may be detected at lower energy~\cite{Heurtier:2022rhf}. We aim in the future to enlarge our study to include the discussion of cosmic inflation and how it is affected in the cosmological framework that we have studied in this paper. 


\textit{To conclude,} we emphasize, once again, that hunting for apparently unrelated signals from the laboratory and from the sky will help us to break degeneracies in beyond the SM theories of cosmology that involve multiple energy scales, and provide the community with a powerful way to search for new physics.

\medskip

\section*{Acknowledgement}
AG thanks Konstantinos Dimopoulos and Pedro Schwaller for providing useful feedback on the manuscript. The work of LH is supported in part by the U.K. Science and Technology Facilities Council (STFC) under Grant ST/P001246/1. This work was made possible by Institut Pascal at Université Paris-Saclay during the Paris-Saclay Astroparticle Symposium 2021, with the support of the P2IO Laboratory of Excellence (programme “Investissements d’avenir” ANR-11-IDEX-0003-01 Paris-Saclay and ANR-10-LABX-0038), the P2I research departments of the Paris-Saclay university, as well as IJCLab, CEA, IPhT, APPEC, the IN2P3 master projet UCMN and EuCAPT.

\bibliographystyle{apsrev4-1}
\bibliography{main.bib}


\newpage

\end{document}